\documentclass[sigconf, screen]{acmart}
\usepackage{booktabs}
\usepackage{multirow}
\usepackage{tabularx}
\usepackage{xspace}
\usepackage{csquotes}
\usepackage{float}
\usepackage{hyperref}

\AtBeginDocument{%
  }

\setcopyright{acmcopyright}
\acmConference{...}
\acmBooktitle{...}
\acmPrice{...}
\acmDOI{...}
\acmISBN{...}

\renewcommand\footnotetextcopyrightpermission[1]{}

\begin{document}

\title{Constraint-Aware Conversational Job Recommendation in Code-Mixed Low-Resource Settings}

\author{Md Arman Hossain}
\authornote{Both authors contributed equally to this research.}
\affiliation{%
  \institution{East West University}
  \city{Dhaka}
  \country{Bangladesh}}
\email{2022-2-60-043@std.ewubd.edu}

\author{Mubashir Jawad}
\authornotemark[1]
\affiliation{%
  \institution{East West University}
  \city{Dhaka}
  \country{Bangladesh}}
\email{2022-2-60-039@std.ewubd.edu}

\author{Fariha Khandaker Moon}
\affiliation{%
  \institution{East West University}
  \city{Dhaka}
  \country{Bangladesh}}
\email{2022-2-60-047@std.ewubd.edu}

\author{Sonia Binte Siraj}
\affiliation{%
 \institution{East West University}
 \city{Dhaka}
 \country{Bangladesh}}
 \email{2022-2-60-087@std.ewubd.edu}

\author{Masfiqur Rahaman}
\affiliation{%
  \institution{University of California San Diego}
  \city{San Diego, CA}
  \country{USA}}
\email{marahaman@ucsd.edu}

\author{Raihan ul Islam}
\affiliation{%
  \institution{East West University}
  \city{Dhaka}
  \country{Bangladesh}}
\email{raihan.islam@ewubd.edu}

\author{Ahmed Wasif Reza}
\affiliation{%
  \institution{East West University}
  \city{Dhaka}
  \country{Bangladesh}}
\email{wasif@ewubd.edu}

\author{Nafis Sadeq}
\correspondingauthor
\affiliation{%
  \institution{East West University}
  \city{Dhaka}
  \country{Bangladesh}}
\email{nafis.sadeq@ewubd.edu}

\renewcommand{\shortauthors}{Hossain et al.}

\begin{abstract}
Conversational job recommendation requires jointly modeling semantic relevance, user preferences, eligibility requirements, and the noisy language used in real-world career discussions. These challenges are especially pronounced in low-resource, code-mixed settings, where strict constraint matching can incorrectly eliminate otherwise suitable jobs. We introduce JobCCC, a conversational job recommendation benchmark for Bangladesh comprising 22,410 structured job postings and 988 multi-turn career-advice dialogues derived from regional Reddit communities. Each dialogue is annotated with evolving seeker preferences and linked to a ground-truth job, and is evaluated in semantically equivalent English and Romanized Bangla--English (Banglish) variants. We compare sparse BM25 retrieval, multilingual dense retrieval, and their hard-constraint-filtered counterparts against Weighted Soft-Constraint-Aware Ranking (W-SCAR), our multi-criteria ranking framework that combines lexical relevance, semantic relevance, and graded utilities for experience, location, education, and salary using the Technique for Order Preference by Similarity to Ideal Solution (TOPSIS). Experiments reveal that strict filtering consistently degrades retrieval because incomplete extraction and brittle attribute matching irreversibly remove relevant jobs. W-SCAR avoids destructive pruning and achieves more balanced performance across the two language conditions, obtaining 37.37\% and 38.43\% Hit@10 on English and Banglish, respectively. The results further expose complementary cross-lingual behavior: dense retrieval performs best on English, whereas BM25 benefits substantially from lexical anchors in Romanized Banglish. These findings demonstrate that conversational job recommendation in noisy multilingual environments is better formulated as soft-constraint multi-criteria ranking than as relevance retrieval followed by binary eligibility filtering. The code and dataset are publicly available at \href{https://github.com/M-Jawad01/Conversational-Job-Recommendation-System-LLM}{GitHub} and \href{https://huggingface.co/datasets/Armans33115/JobCCC-Conversational-Job-Recommendation-Bangladesh}{Hugging Face}, respectively.
\end{abstract}

\maketitle
\section{Introduction}
Conversational recommender systems (CRS) support item discovery through natural-language interaction. Unlike conventional recommender systems that rely primarily on static profiles or historical interactions, a CRS can progressively infer user needs from multi-turn dialogue and revise its recommendations as additional preferences and constraints are expressed~\cite{li2018neurips,Fu2020recsys}. Existing CRS research has largely focused on domains such as movies~\cite{chen-2019-kbrd,zhou-2020-kdd-kgsf}, music~\cite{jin-2019-cikm,cai-2021-iui}, restaurants~\cite{Christakopoulou-2016-kdd,bursztyn-etal-2021-doesnt}, and e-commerce products~\cite{sun-zhang-2018-sigir,liu-2023-sigir}, where the primary objective is often to identify items that align with subjective preferences. Job recommendation presents a substantially different problem because recommendation quality depends not only on preference matching, but also on candidate eligibility, occupational requirements, and the temporal validity of job postings.

Traditional job recommendation systems commonly use content-based matching between candidate profiles and job descriptions \cite{Alsaif2022_21}, graph-based collaborative models \cite{ShalabyWalid2017_05}, or combinations of structured and textual features. Although these approaches are scalable, they generally treat candidate qualifications and job requirements as fixed profile attributes and provide limited support for preferences that emerge incrementally during a conversation \cite{LeRan2019_07}. This limitation is consequential because job seekers frequently express incomplete, evolving, or negotiable requirements. For example, a seeker may initially request an entry-level software position in a specific location, later indicate willingness to relocate, and subsequently specify a minimum salary or a preference for remote work. A conversational system must therefore integrate preferences and constraints distributed across multiple turns rather than match a job against a single static query.

Job recommendation also differs from recommendation in lower-stakes domains in several important respects. First, unsuitable recommendations may affect a seeker's income, mobility, safety, and longer-term career trajectory. Second, job matching involves eligibility conditions such as education, experience, location, age range, and occasionally employer-specified demographic requirements. Third, the system must account for softer considerations, including salary expectations, industry preferences, work environment, career development, field-work requirements, and tolerance for sales targets. Fourth, job postings are temporally bounded, making application deadlines an important component of recommendation validity. Finally, real-world job metadata is frequently incomplete, inconsistently structured, or linguistically heterogeneous \cite{LeRan2019_07}. These difficulties are amplified in low-resource settings, where job descriptions and user queries may combine English, Bangla, Romanized Bangla, local abbreviations, and domain-specific terminology \cite{ChakravarthiBharathiRaja2020_11,MishraHarsh2025_32}.

In this work, we study conversational job recommendation for Bangladesh in a seeker--expert setting. A job seeker communicates qualifications, preferences, constraints, and career concerns through a multi-turn dialogue, while the system must retrieve suitable jobs from a large and heterogeneous collection of postings. Our setting is motivated by the Bangladeshi labor market, where conversational queries may include standard English, Romanized Bangla, and Bangla--English code-mixing. Job postings may similarly combine structured metadata with free-text descriptions, English technical terms, local job titles, and abbreviations such as `SR' for Sales Representative or `FMCG' for fast-moving consumer goods. These characteristics create a retrieval problem in which semantic relevance, lexical overlap, structured compatibility, and linguistic variation must be considered jointly.

A central challenge in this setting concerns the treatment of user constraints. A common strategy is to interpret extracted preferences as hard predicates and remove every posting that fails at least one requirement. Such filtering is attractive because it appears to reduce the search space and prevent infeasible recommendations. However, it assumes that dialogue-level preference extraction is accurate, that conversational phrases correspond directly to normalized database fields, and that every expressed preference should be treated as inviolable. These assumptions are often unrealistic. A free-text education preference such as `degree in accounting,' for example, may not match a structured field containing `Bachelor,' even though the two descriptions are compatible. Similarly, missing or unnormalized salary and experience fields may cause a suitable posting to be rejected. Once the ground-truth job has been removed by binary filtering, no subsequent retrieval model can recover it.

We therefore investigate whether conversational job recommendation is better formulated as a soft-constraint ranking problem. Rather than eliminating candidates, our proposed Weighted Soft-Constraint-Aware Ranking framework, W-SCAR, assigns each posting graded utilities for experience, location, education, and salary compatibility. These utilities are combined with sparse lexical relevance and dense semantic relevance through a multi-criteria decision-making procedure. Criterion weights are determined using the Analytic Hierarchy Process, and candidates are ranked using the Technique for Order of Preference by Similarity to Ideal Solution. This formulation preserves the influence of user constraints while allowing a strong semantic or lexical match to remain recoverable under moderate preference mismatches, incomplete metadata, or extraction errors.

Table~\ref{tab:job_info} presents an example of the heterogeneous job metadata considered in this work. The posting concerns a Sales Representative position and combines English metadata with Bangla free-text fields. To make the multilingual content accessible to English readers, the corresponding English translations are provided alongside the original Bangla text. The example also illustrates several practical challenges: benefits are underspecified, the location is broad, the education requirement is nonrestrictive, and the responsibilities imply sales targets and customer-facing work that may not be represented in a normalized field.

\begin{table}[t]
\centering
\caption{Example job metadata containing structured fields, Bangla text with English translations, English skill terms, local abbreviations, and partially specified information.}
\label{tab:job_info}
\small
\begin{tabular}{p{0.15\linewidth}p{0.80\linewidth}}
\toprule
\textbf{Field} & \textbf{Example Value} \\
\midrule
Job title & Sales Representative (SR) \\
Category & Sales Rep (SR) \\
Company & Eureka Food and Beverage \\
Description &
Responsibilities and duties include selling the company's products, developing new markets, improving relationships with customers, and achieving sales targets (Translated)\\
Skills & Market Research, Marketing, Sales and Marketing, Sales Target, Customer Service \\
Salary & 18,000--22,000 BDT \\
Location & Anywhere in Bangladesh; company address in Ashulia, Savar, Dhaka \\
Education & Any \\
Experience & 2--5 years \\
Deadline & January 5, 2026 \\
Benefits &
Negotiable (Translated)\\
Instruction &
Interested candidates are encouraged to apply (Translated)\\
\bottomrule
\end{tabular}
\end{table}

A generic relevance model may rank this posting highly for any seeker who mentions sales. A job-specific conversational recommender must additionally determine whether the experience requirement is compatible with the seeker's profile, whether the salary range is acceptable, whether the location is feasible, and whether the role's field-sales and target-oriented responsibilities align with the seeker's preferences \cite{ParameswaranAditya2011_24,MathiasUta2021_06,QixinWang2025_08}. The model must perform this reasoning under noisy conversational inputs and partially structured job records. It must also remain robust when semantically equivalent requests are expressed in standard English or Romanized Bangla.

To study these challenges, we construct a benchmark comprising 22,410 job postings and 988 annotated multi-turn career-advice dialogues. Each dialogue is associated with structured seeker preferences and a ground-truth job posting. We synthesize semantically equivalent English and Banglish variants for collected ground truth dialogues, allowing the effect of linguistic representation to be examined while holding the underlying intent, constraints, and relevance labels constant. The experimental comparison includes BM25, multilingual dense retrieval, hard-constraint variants of both retrievers, and the proposed W-SCAR framework. The results show that hard filtering consistently reduces retrieval effectiveness because incomplete extraction and brittle attribute matching remove relevant postings before ranking. They also reveal complementary cross-lingual behavior: dense retrieval performs substantially better on English, whereas sparse retrieval benefits from lexical anchors in Romanized Banglish. W-SCAR avoids destructive pruning and produces substantially more balanced performance across the two language variants.

We investigate three research questions: \textbf{RQ1}, how effectively sparse, dense, and hard-constraint ranking methods retrieve relevant jobs from multi-turn dialogues; \textbf{RQ2}, how graded soft-constraint ranking method performs compared with unfiltered and hard-constraint baselines; and \textbf{RQ3}, how these methods perform across semantically equivalent English and Romanized Bangla queries. To address these questions, we make the following contributions:

\begin{enumerate}
\item We construct a benchmark containing 22,410 structured job postings and 988 annotated career-advice dialogues, with semantically equivalent English and Romanized Bangla variants linked to the same preference annotations and ground-truth jobs.

\item We introduce W-SCAR, a soft-constraint multi-criteria ranking framework that combines BM25 relevance, multilingual dense similarity, and graded utilities for experience, location, education, and salary compatibility using Analytic Hierarchy Process (AHP) based weighting and TOPSIS ranking.

\item Through controlled experiments, we show that hard filtering consistently reduces retrieval effectiveness by prematurely removing relevant jobs, while sparse and dense retrieval exhibit complementary cross-lingual behavior and W-SCAR provides more balanced performance across the two language conditions.
\end{enumerate}

\section{Related Work}

\subsection{Job Recommendation}
Early job recommendation systems formulated recommendation as a profile matching problem between job seekers and job descriptions. Traditional approaches primarily relied on keyword extraction from unstructured job postings, clustering techniques, and historical user interactions~\cite{Mhamdi2020_30}. Subsequent work incorporated natural language signals to model semantic similarity between resumes and job descriptions \cite{Alsaif2022_21}. As recruitment platforms expanded, graph-based methods became increasingly popular for modeling user--job interactions while alleviating data sparsity and cold-start issues \cite{ShalabyWalid2017_05}. More recently, several studies have explored converting structured interaction graphs into textual representations to enable reasoning with Large Language Models (LLMs) \cite{LikangWu2023_22}.

Despite these advances, job recommendation differs from traditional recommendation tasks because successful matching must satisfy the preferences of both employers and job seekers. Consequently, recommendation quality depends not only on semantic relevance but also on satisfying hard constraints such as salary requirements, geographic location, and working hours \cite{LeRan2019_07,MathiasUta2021_06,ParameswaranAditya2011_24}. Existing recommendation models often optimize similarity without explicitly enforcing such constraints, resulting in infeasible recommendations.

Recent studies have therefore incorporated constraint-aware reasoning into job recommendation. Case-based reasoning frameworks have been proposed to jointly model soft and hard constraints while providing interpretable recommendations \cite{DavronjonShodmonov2026_14}. In parallel, hybrid lexical-semantic retrieval methods improve candidate retrieval by combining exact term matching with semantic similarity \cite{HusseinAlAwad2026_33}. However, these systems remain largely static and cannot iteratively refine user preferences through interaction.

\subsection{Conversational Recommender Systems}

CRS address the limitations of static recommendation by eliciting user preferences through multi-turn dialogue \cite{LeiWenqiang2020_01,sun-zhang-2018-sigir}. Early CRS models demonstrated the effectiveness of interactive preference elicitation but often struggled with complex user intents and negative feedback \cite{Jannach2021_28}. These challenges are particularly important in recruitment, where both recruiters and job seekers typically refine their preferences during the recommendation process \cite{MentecFrancois2021_09}. Although task-oriented dialogue systems have achieved promising results in constrained domains such as e-commerce \cite{YanZhao2017_27}, applying them to recruitment requires substantially richer contextual reasoning \cite{QixinWang2025_08}.

Recent advances in LLMs have significantly expanded the capabilities of conversational recommendation \cite{LikangWu2024_02,WayneXinZhao2026_03}. LLMs enable recommendation to be formulated as a language understanding and reasoning task rather than relying solely on interaction matrices \cite{ShijieGeng2023_25}. Prior studies demonstrate strong performance in explanation generation, conversational reasoning, and zero-shot recommendation \cite{ZhaoZihuai2024_20,YasharDeldjoo2024_29,He2023_23,YupengHou2024_19}. To further improve reliability, recent work has investigated chain-of-thought reasoning, reflective recommendation, multi-agent collaboration, and retrieval-augmented dialogue grounding \cite{YiFang2025_13,LeiWenqiang2020_18,HuiZheng2026_12,JiaoChen2025_15,ZhuYaochen2025_10}. Nevertheless, these systems are predominantly evaluated on clean, high-resource datasets and do not explicitly consider noisy, code-mixed user interactions.

\subsection{Retrieval in Code-Mixed Settings}

Information retrieval remains a critical component of conversational recommendation because the quality of downstream reasoning depends on the relevance of retrieved candidates. However, retrieval in low-resource and code-mixed languages remains challenging \cite{Karthikeyan2025_31}. On many online platforms, users communicate using Romanized, code-mixed text such as Banglish, introducing spelling variation, transliteration ambiguity, and inconsistent language usage. The scarcity of annotated corpora and linguistic resources further complicates retrieval in these settings \cite{ChakravarthiBharathiRaja2020_11}.

Existing lexical retrieval methods are sensitive to transliteration and spelling variations, whereas dense retrieval models often struggle to align multilingual representations under code-mixed inputs \cite{MishraHarsh2025_32,DoJunggeun2024_04}. Moreover, benchmark studies have shown that retrieval performance on clean datasets does not necessarily generalize to noisy, zero-shot scenarios \cite{NandanThakur2021_16}. Consequently, neither sparse nor dense retrieval models consistently support conversational recommendation over code-mixed user queries.

Existing work has studied code-mixed retrieval and conversational recommendation largely as separate problems. However, limited attention has been given to how structured user constraints should be incorporated when preferences are expressed through noisy, code-mixed, multi-turn dialogue. In particular, hard filtering may be brittle when extracted preferences and job attributes are incomplete or lexically mismatched. Our work addresses this gap by jointly evaluating sparse and dense retrieval under English and Banglish queries and by introducing W-SCAR, which integrates lexical relevance, semantic relevance, and graded constraint compatibility within a unified ranking framework.

\section{Dataset}
\label{sec:dataset}
JobCCC contains 22,410 job listings and 988 multi-turn seeker--expert career conversations. The two components are linked through job identifiers. Each dialogue is available in both code-mixed Banglish and standard English, with utterance-level alignment between the two versions.

\subsection{Job Details Construction}
\textbf{Collection and preprocessing.}
Job listings were collected from BDJobs.com, a major online job portal in Bangladesh, in accordance with the website's \texttt{robots.txt}. The initial collection contained more than 24,500 raw JSON records. We removed residual HTML, normalized whitespace, and corrected recurring encoding errors through targeted string substitutions. Approximately 8\% of the records were discarded because they were duplicated, expired, or contained fewer than 50 words. Personally identifiable information (PII) and highly sparse administrative fields, such as application URLs with approximately 98\% missing values, were also removed.

Several unstructured attributes were converted into structured features. Age and experience requirements were parsed using regular expressions to obtain minimum and maximum values. Location information was normalized into district-level and fine-grained location fields, while educational requirements were mapped to education levels and fields of study using keyword-based rules. We further corrected 1,179 category assignments based on job-content keywords. After preprocessing, the dataset contains 22,410 listings across 45 job categories.

\textbf{Semantic enrichment.}
Job descriptions are typically long and unstructured, averaging approximately 300--500 words. We therefore used a large language model to extract six semantic attributes: technical skills, soft skills, responsibility summaries, language requirements, benefits, and application mode. The generated outputs were constrained to structured JSON. A final feature-selection step removed the low-value administrative fields, including application instructions, company URLs, internal category identifiers, and application-control fields.

The resulting job dataset contains 40 attributes. Representative fields include job\_id, company\_id, category\_id, company\_name, job\_title, work\_place, job\_benefits, deadline, vacancies, skills, gender, salary, district, fine\_grained\_location, education\_level, education\_subject, age, experience, and responsibilities\_summary.

\begin{table}
\centering
\caption{Summary of job-market dataset characteristics.}
\label{tab:job_market_stats}
\small
\setlength{\tabcolsep}{4pt}
\renewcommand{\arraystretch}{0.95}
\begin{tabularx}{\columnwidth}{@{}p{0.36\columnwidth}X@{}}
\toprule
\textbf{Category} & \textbf{Distribution} \\
\midrule
Educational requirements &
Bachelor's degree (85.3\%); Diploma (13.7\%); Master's degree (1.0\%) \\
Academic domains &
Computer Science (41.4\%); Business Administration (32.0\%);
Engineering (13.7\%); Science/Engineering (12.9\%) \\
Experience requirements &
Fresher (17.47\%); 1 year (23.24\%); 2 years (22.51\%); 3 years (17.48\%);
5 years (13.42\%) \\
Gender preference &
Both/neutral (67.1\%); Male (26.7\%); Female (6.2\%) \\
Geographic distribution &
Dhaka (56.9\%); Anywhere in Bangladesh (14.6\%); Gazipur (6.1\%) \\
Industry categories &
General Management/Administration (25.3\%);
Marketing/Sales (19.9\%); IT \& Telecom (8.5\%) \\
\bottomrule
\end{tabularx}
\end{table}

\subsection{Seeker--Expert Dialogues}
The conversational component was constructed from employment-related discussions collected from regional Reddit communities, including r/Dhaka and r/Bangladesh. The initial collection contained approximately 74,000 submissions and comments. An LLM-based relevance filter retained 14,000 career-related interactions, which were linked to their parent posts to reconstruct approximately 3,000 conversational threads.

After programmatic validation and filtering, 1,010 seed posts were retained and used to construct candidate dialogues. Additional quality control, annotation, and multilingual processing resulted in a final set of 988 dialogues. Each dialogue is represented in two aligned linguistic forms: standard English and Banglish. The final corpus contains approximately 9,000 conversational turns, with dialogues containing 5--12 replies and an average of 8.15 replies per thread.

To connect conversational preferences with the job collection, each dialogue was annotated with evolving job-seeking constraints such as preferred role, location, experience level, employment type, and skills. At relevant turns, these constraints are associated with a ground-truth set of matching jobs from the 22,410 listings. All of the 988 dialogues contain at least one turn with a non-empty ground-truth job set. The resulting dialogue-to-job annotations provide direct supervision for evaluating whether a system can infer user requirements from multi-turn, code-mixed conversations and retrieve appropriate job listings.

\subsection{Dataset Analysis}

The job collection  captures several constraints relevant to recommendation. Dhaka accounts for 56.6\% of listings, 95.2\% of positions are full-time, and 32.9\% specify a gender preference, including 26.7\% male and 6.2\% female preferences.

Similar constraints appear in the conversational data. Frequent location-related expressions include ``Bangladesh'' (1,928 mentions) and ``Dhaka'' (965), while common employment and career terms include ``data analysis'' (705), ``fresher'' (618), ``entry-level'' (580), ``part-time'' (539), and ``remote'' (506). This overlap between conversational intent and structured job attributes enables evaluation of constraint-aware job retrieval from natural, multi-turn user interactions.

\begin{table}[htbp]
\centering
\caption{Top 10 in-demand job roles }
\small
\begin{tabular}{l l r}
\toprule
\textbf{Rank} & \textbf{Job Title} & \textbf{Frequency} \\
\midrule
1 & Sales Executive & 492 \\
2 & Assistant Teacher & 488 \\
3 & Marketing Executive & 371 \\
4 & Graphic Designer & 357 \\
5 & Accountant & 301 \\
6 & Digital Marketing Manager & 266 \\
7 & Front Desk Executive & 214 \\
8 & Medical Officer & 195 \\
9 & Sales Officer & 192 \\
10 & Driver & 161 \\
\bottomrule
\end{tabular}
\label{tab:top10_jobs}
\end{table}

\begin{table}[htbp]
\centering
\caption{Salary distribution in the local job market}
\small
\begin{tabular}{l c r}
\toprule
\textbf{Salary Range (BDT)} & \textbf{Number of Jobs} & \textbf{Percentage (\%)} \\
\midrule
$<$ 15,000 & 984 & 15.91 \\
15,000 - 25,000 & 2126 & 34.38 \\
25,000 - 35,000 & 1232 & 19.93 \\
35,000 - 50,000 & 805 & 13.02 \\
50,000 - 75,000 & 570 & 9.22 \\
75,000 - 1,00,000 & 202 & 3.27 \\
$>$ 1,00,000 & 264 & 4.27 \\
\bottomrule
\end{tabular}
\label{tab:salary_dist}
\end{table}

Table~\ref{tab:top10_jobs} shows the top 10 most in-demand jobs in JobCCC based on how often they appear. The list is led by \textit{Sales Executive} (492 postings) and \textit{Assistant Teacher} (488 postings).There is a high demand for entry-level and customer-facing jobs like sales and marketing. The table also shows a good mix of different job types from corporate office roles like \textit{Accountant} to service-oriented jobs like \textit{Front Desk Executive} and \textit{Driver}. 
Table~\ref{tab:salary_dist} shows the salary distribution over 54\% of the jobs pay between 15,000 and 35,000 BDT, which matches the high number of entry-level roles in the market. In contrast, only 7.5\% of jobs offer more than 75,000 BDT.

\begin{table}[htbp]
\centering
\caption{Top 10 districts for job opportunities}
\small
\begin{tabular}{l r}
\toprule
\textbf{District} & \textbf{Number of Jobs} \\
\midrule
Dhaka & 12756 \\
Anywhere in Bangladesh & 3274 \\
Gazipur & 1364 \\
Chattogram & 1026 \\
Narayanganj & 670 \\
Cox's Bazar & 450 \\
Sylhet & 221 \\
Mymensingh & 176 \\
Cumilla & 153 \\
Khulna & 150 \\
\bottomrule
\end{tabular}
\label{tab:top_districts}
\end{table}

\begin{table}[htbp]
\centering
\caption{Top 15 employment hubs within Dhaka, excluding job postings where a fine-grained location within Dhaka was not mentioned.}
\small 
\begin{tabular}{l r}
\toprule
\textbf{Area in Dhaka} & \textbf{Number of Jobs} \\
\midrule
Uttara & 1010  \\
Gulshan & 966  \\
Mirpur & 617  \\
Banani & 568  \\
Savar & 482  \\
Dhanmondi & 370  \\
Mohakhali & 284  \\
Ashulia & 272  \\
Baridhara & 237  \\
Mohammadpur & 185  \\
Tejgaon & 179 \\
Motijheel & 152 \\
Badda & 129  \\
Basundhara R/A & 113  \\
Banasree & 106  \\
\bottomrule
\end{tabular}
\label{tab:top_dhaka_areas}
\end{table}

Tables~\ref{tab:top_districts} and~\ref{tab:top_dhaka_areas}  highlight Dhaka's overwhelming dominance in the job market with 12,756 listings. Interestingly, "Anywhere in Bangladesh" ranks second (3,274 postings) .It  indicates a strong demand for remote or flexible roles while major districts like Gazipur (1,364) and Chattogram (1,026) trail significantly behind.
A large portion of listings (5,440) do not specify an exact neighborhood within Dhaka. Commercial zones like Uttara, Gulshan, Mirpur, and Banani offer the most opportunities. This heavy centralization underscores why geographic location remains a critical constraint in this system.

\begin{table}[htbp]
\centering
\caption{Distribution of Required Experience Levels in Job Postings}
\small
\begin{tabular}{l c r}
\toprule
\textbf{Experience Level} & \textbf{Number of Vacancies} & \textbf{Percentage (\%)} \\
\midrule
Fresher (0 Years) & 3935 & 17.47 \\
Junior (1-2 Years) & 8505 & 37.76 \\
Mid-Level (3-5 Years) & 6555 & 29.10 \\
Senior (6-8 Years) & 1703 & 7.56 \\
Expert (9+ Years) & 1827 & 8.11 \\
\bottomrule
\end{tabular}
\label{tab:experience_dist}
\end{table}

From Table~\ref{tab:experience_dist}, a massive 84\% of the vacancies are targeted at early-career professionals, with Junior (1-2 years) and Mid-Level (3-5 years) roles making up the vast majority. Freshers also have a strong presence, accounting for over 17\% of the jobs. In contrast, senior and expert roles requiring 6 or more years of experience are relatively scarce. This clear market bias towards early-career candidates perfectly matches our conversational data, where a large number of users specifically searched for "fresher" and "entry-level" positions.

\begin{table}[htbp]
\centering
\small %
\caption{Linguistic Breakdown of Job Postings in the Dataset}
\begin{tabular}{p{3cm} r r}
\toprule
\textbf{Linguistic Category} & \textbf{Number of Vacancies} & \textbf{Percentage (\%)} \\
\midrule
English & 20569 & 91.34 \\
Bengali & 1712 & 7.60 \\
Code-Mixed (English + Bengali) & 237 & 1.05 \\
\bottomrule
\end{tabular}
\label{tab:linguistic_dist}
\end{table}

Table~\ref{tab:linguistic_dist}  shows the two components exhibit substantial linguistic and contextual variation. Among job listings, 91.34\% are written entirely in English, 1.05\% contain English--Bangla code-mixing, and 7.60\% are written primarily in Bangla. Language use also varies across job categories: 98.4\% of IT listings are in English, whereas some non-technical categories contain up to 13.7\% code-mixed and 22.5\% Bangla-only postings. These distributions motivate evaluation under multilingual and code-mixed settings.

\section{Methodology}
\label{sec:method}

We propose W-SCAR, a multi-criteria ranking framework for conversational job recommendation. W-SCAR jointly considers textual relevance and structured preference compatibility without eliminating candidates through hard filtering. Instead, user preferences are represented as graded utilities and combined with lexical and semantic relevance using AHP-weighted TOPSIS.

\begin{figure}[H]
    \centering
    \includegraphics[width=1\linewidth]{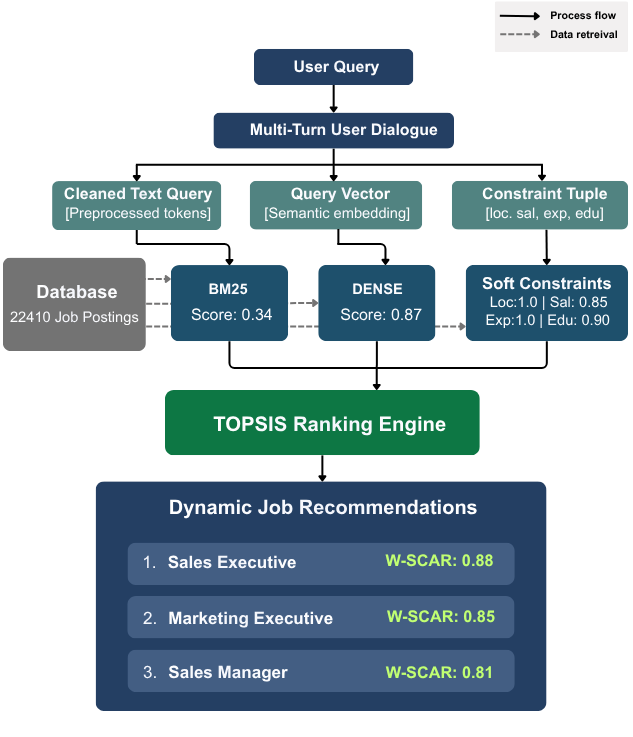}
    \caption{Architecture of the Proposed W-SCAR Framework}
    \label{fig:placeholder}
\end{figure}

\subsection{Problem Formulation}

Let $\mathcal{J} = \{d_1,\ldots,d_N\}$ denote a corpus of $N$ job postings. Each posting $d_j$ contains unstructured text and structured attributes including location, education requirement, salary range $[s_j^{\min}, s_j^{\max}]$, and experience range $[e_j^{\min}, e_j^{\max}]$.

A conversational query is represented by a multi-turn thread
\begin{equation}
T = (p,c_1,\ldots,c_m),
\end{equation}
where $p$ is the initial seeker post and $c_1,\ldots,c_m$ are subsequent turns. Each thread is associated with a ground-truth job posting $g_T \in \mathcal{J}$. The objective is to produce a ranked list $R_T$ in which $g_T$ is ranked as highly as possible.

Two types of information are derived from each thread. The dialogue text $q_T$ is formed by concatenating the initial post and subsequent turns. Structured preferences are represented as
\begin{equation}
\mathbf{c}_T =
\left(
c_T^{\mathrm{loc}},
c_T^{\mathrm{edu}},
c_T^{\mathrm{sal}},
c_T^{\mathrm{exp}}
\right).
\end{equation}
Preferences are obtained from the turn-level annotations by selecting the most recently stated value for each attribute, with the initial post used when no corresponding preference appears in later turns. A preference may remain unspecified if it cannot be recovered from the dialogue.

Each conversation is available in semantically equivalent English and Banglish variants. Both variants share the same structured preferences and ground-truth job, allowing the effect of query language to be evaluated while keeping the recommendation target fixed.

\subsection{Textual Relevance}

W-SCAR uses complementary sparse and dense relevance signals. Each job posting is represented by text combining its title, required skills, technical skills, and full description.

For sparse relevance, the dialogue and job text are normalized and tokenized, and relevance is computed using Okapi BM25~\cite{robertson2009bm25}:
\begin{equation}
r_j^{\mathrm{lex}} = \mathrm{BM25}(q_T,d_j).
\end{equation}

For semantic relevance, the dialogue is encoded using an instruction-tuned multilingual E5 model~\cite{wang2024multilinguale5textembeddings}, while job embeddings are precomputed from the same textual representation. Dense relevance is measured using cosine similarity:
\begin{equation}
r_j^{\mathrm{sem}}
=
\max\left(0,\cos(\mathbf{q},\mathbf{d}_j)\right),
\end{equation}
where $\mathbf{q}$ and $\mathbf{d}_j$ denote the query and job embeddings, respectively.

\subsection{Soft-Constraint Utilities}

W-SCAR represents each structured preference as a bounded utility rather than a binary constraint. A satisfied preference receives maximum utility, while mismatches receive graded penalties. Unspecified preferences are assigned neutral utility so that missing information does not eliminate a candidate.

\paragraph{Salary.}
Let $r$ denote the requested salary extracted from the conversation. The salary utility is
\begin{equation}
u_j^{\mathrm{sal}} =
\begin{cases}
1.0, & s_j^{\min} \le r \le s_j^{\max}, \\
0.85, & s_j^{\min} = s_j^{\max} = 0, \\
\max\left(0.4,\frac{s_j^{\max}}{r}\right), & s_j^{\max} < r, \\
1.0, & \text{otherwise}.
\end{cases}
\end{equation}
Thus, jobs below the requested salary are penalized according to the magnitude of the shortfall but remain eligible for recommendation.

\paragraph{Experience.}
For a numerical experience preference $r$,
\begin{equation}
u_j^{\mathrm{exp}} =
\begin{cases}
1.0, & r \ge e_j^{\min}, \\
\max\left(0.3,\frac{r}{e_j^{\min}}\right), & r < e_j^{\min}.
\end{cases}
\end{equation}
Entry-level preferences receive utility $1.0$ when the minimum required experience is at most one year and $0.4$ otherwise.

\paragraph{Location.}
A bidirectional substring match between the preferred and advertised locations receives utility $1.0$. Location-independent requirements such as ``remote,'' ``anywhere,'' or unspecified locations receive $0.95$, while other mismatches receive $0.4$.

\paragraph{Education.}
A bidirectional substring match receives utility $1.0$, an unspecified education requirement receives $0.9$, and other mismatches receive $0.5$. The moderate mismatch penalty accounts for variation between free-text educational preferences and structured job requirements.

\subsection{Multi-Criteria Ranking}

For each query, W-SCAR represents every job posting using six criteria:
\begin{equation}
\mathbf{x}_j =
\left(
r_j^{\mathrm{sem}},
r_j^{\mathrm{lex}},
u_j^{\mathrm{exp}},
u_j^{\mathrm{loc}},
u_j^{\mathrm{edu}},
u_j^{\mathrm{sal}}
\right).
\end{equation}
These vectors form a decision matrix $X \in \mathbb{R}^{N \times 6}$.

The relative importance of the six criteria is determined using AHP. Pairwise judgments prioritize semantic relevance and experience compatibility, while lexical relevance, location, education, and salary provide complementary signals. The resulting normalized weight vector is
\begin{equation}
\mathbf{w}
=
(0.367,\;0.137,\;0.232,\;0.128,\;0.081,\;0.055),
\end{equation}
corresponding to semantic relevance, lexical relevance, experience, location, education, and salary, respectively. The AHP consistency ratio is $0.0112$, below the conventional threshold of $0.1$.

Because the six criteria have different numerical ranges, each criterion is
scaled to $[0,1]$ using min--max normalization:
\begin{equation}
\tilde{x}_{j,c}
=
\frac{x_{j,c}-\min_{j'}x_{j',c}}
{\max_{j'}x_{j',c}-\min_{j'}x_{j',c}},
\end{equation}
where a unit denominator is used when a criterion is constant across all
candidates. The normalized criteria are then weighted as
\begin{equation}
v_{j,c}=w_c\tilde{x}_{j,c}.
\end{equation}

TOPSIS ranks jobs according to their proximity to the ideal combination of criteria. Since all criteria are benefit criteria, the positive and negative ideal values are
\begin{equation}
A_c^{+} = \max_j v_{j,c},
\qquad
A_c^{-} = \min_j v_{j,c}.
\end{equation}

For each posting, the distances from the positive and negative ideals are
\begin{equation}
d_j^{+}
=
\left(
\sum_{c=1}^{6}
\left(v_{j,c} - A_c^{+}\right)^2
\right)^{1/2},
\end{equation}
and
\begin{equation}
d_j^{-}
=
\left(
\sum_{c=1}^{6}
\left(v_{j,c} - A_c^{-}\right)^2
\right)^{1/2}.
\end{equation}

The final W-SCAR score is the TOPSIS closeness coefficient
\begin{equation}
C_j
=
\frac{d_j^{-}}{d_j^{+} + d_j^{-}},
\end{equation}
where a larger value indicates greater proximity to the ideal relevance and preference profile. Jobs are ranked in descending order of $C_j$, and the highest-ranked postings are returned as recommendations.

W-SCAR therefore preserves the full candidate set and represents preference violations as graded penalties rather than irreversible exclusions. This formulation is suitable for conversational recommendation settings in which preferences may be incomplete, inconsistently expressed, or imperfectly extracted.

\section{Experiment}

\subsection{Experimental Setup}
\label{sec:experimental_setup}

\paragraph{\textbf{Evaluation data.}}
We evaluate all methods on the 988 annotated multi-turn dialogues in JobCCC and the corresponding corpus of 22,410 job postings. Each dialogue represents a career-advice interaction in which the seeker expresses qualifications, preferences, and constraints over multiple turns, and is associated with a single ground-truth job posting used for retrieval evaluation. Each dialogue is available in semantically equivalent English and Banglish variants. The paired variants share the same structured preference annotations and ground-truth job, allowing us to evaluate the effect of linguistic representation while holding the underlying recommendation target fixed.

\paragraph{\textbf{Baselines.}}
We consider two standard retrieval paradigms: sparse lexical retrieval and multilingual dense retrieval. For both methods, the query consists of the concatenated multi-turn dialogue, while each job is represented using its title, required skills, technical skills, and description.

For sparse retrieval, we use Okapi BM25. Dialogue and job text are lowercased, punctuation is removed, and whitespace tokenization is applied before indexing and retrieval. For dense retrieval, we use the same instruction-tuned multilingual E5 encoder described in Section~\ref{sec:method}. Job representations are pre-computed, and each dialogue is encoded into the same embedding space at inference time. Candidate jobs are ranked by cosine similarity between the dialogue and job embeddings.

Each retriever is evaluated under two candidate-selection strategies. In the \emph{unfiltered} setting, all 22,410 postings are ranked directly. In the \emph{hard-constraint} setting, the candidate set is first restricted according to four structured seeker preferences: location, education, salary, and experience. Categorical attributes are matched using normalized partial-text matching, while numerical requirements are evaluated using range compatibility. A job is retained only when it satisfies every specified preference. Unspecified preferences do not restrict the candidate set.

Because the English and Banglish versions of a dialogue share the same structured preference annotations, hard-constraint filtering produces the same candidate set for both linguistic variants. For 143 of the 988 dialogues (14.6\%), the conjunction of the extracted constraints removes every candidate. In these cases, we restore the full collection of 22,410 postings before applying the retrieval model. This treatment avoids undefined retrieval outcomes while preserving the effect of hard filtering whenever at least one candidate remains.

The resulting baselines are therefore \textsc{BM25}, \textsc{Filter+BM25}, \textsc{Dense}, and \textsc{Filter+Dense}. We additionally evaluate the proposed W-SCAR framework under the same two language conditions (English + Banglish). Unlike the hard-constraint baselines, W-SCAR does not remove candidates before ranking; instead, it combines BM25 relevance, dense semantic similarity, and graded compatibility scores for experience, location, education, and salary using the TOPSIS formulation described in Section~\ref{sec:method}.

\paragraph{\textbf{Evaluation metrics.}}
Retrieval effectiveness is measured using Hit@10, MRR@10, and NDCG@10. Let $r_i$ denote the rank of the ground-truth job for dialogue $i$. Hit@10 measures the proportion of dialogues for which the ground-truth posting appears among the top ten recommendations:
\begin{equation}
\mathrm{Hit@10}
=
\frac{1}{|\mathcal{Q}|}
\sum_{i \in \mathcal{Q}}
\mathbb{I}[r_i \leq 10].
\end{equation}

MRR@10 additionally rewards methods that place the relevant posting near the top of the ranking:
\begin{equation}
\mathrm{MRR@10}
=
\frac{1}{|\mathcal{Q}|}
\sum_{i \in \mathcal{Q}}
\begin{cases}
1/r_i, & r_i \leq 10,\\
0, & \text{otherwise}.
\end{cases}
\end{equation}

Because each dialogue has a single annotated relevant posting, NDCG@10 reduces to a logarithmically discounted gain determined by the rank of that posting:
\begin{equation}
\mathrm{NDCG@10}
=
\frac{1}{|\mathcal{Q}|}
\sum_{i \in \mathcal{Q}}
\begin{cases}
1/\log_2(r_i+1), & r_i \leq 10,\\
0, & \text{otherwise}.
\end{cases}
\end{equation}

All metrics are macro-averaged over the 988 queries separately for the English and Banglish conditions.

\subsection{Result Analysis}
We organize the result analysis according to the three research questions. Specifically, we examine (RQ1) the retrieval effectiveness of sparse, dense, and hard-constraint ranking methods; (RQ2) the performance of the proposed W-SCAR framework relative to unfiltered and hard-constraint baselines; and (RQ3) the robustness of these methods across semantically equivalent English and Banglish queries. To further explain the observed performance differences, we analyze the effects of hard-constraint filtering, retrieval behavior on the fallback subset, and the characteristics of the underlying lexical and semantic representations. All reported statistics are derived from the query-level experimental results. Each dialogue is associated with a single ground-truth job posting, and a recorded ground-truth rank of zero indicates that the relevant posting was not retrieved.

\begin{table}[htbp]
  \caption{Performance evaluation of W-SCAR.}
  \label{tab:model_performance}
  \centering
  \small
  \begin{tabular}{llccc}
    \toprule
    \textbf{Model} & \textbf{Language} & \textbf{Hit@10} & \textbf{MRR@10} & \textbf{NDCG@10} \\
    \midrule
    \multirow{2}{*}{BM25} 
      & English  & 0.0743  & 0.0343 & 0.0436 \\
      & Banglish & 0.3211 & 0.1262 & 0.1717 \\
    \midrule
    \multirow{2}{*}{Filter+BM25} 
      & English  & 0.0458  & 0.0234 & 0.0287 \\
      & Banglish & 0.2251 & 0.0972 & 0.1248 \\
    \midrule
    \multirow{2}{*}{Dense} 
      & English  & 0.5092 & 0.3501 & 0.3879 \\
      & Banglish & 0.3558 & 0.2193 & 0.2517 \\
    \midrule
    \multirow{2}{*}{Filter+Dense} 
      & English  & 0.2892 & 0.1975 & 0.2192 \\
      & Banglish & 0.1947 & 0.1160 & 0.1346 \\
    \midrule
    \multirow{2}{*}{W-SCAR} 
      & English  & 0.3737 & 0.2406 & 0.2718 \\
      & Banglish & 0.3843 & 0.2340 & 0.2694 \\
    \bottomrule
  \end{tabular}
\end{table}

\paragraph{\textbf{Dense retrieval provides the strongest English baseline.}}
Table~\ref{tab:model_performance} shows substantial differences among the retrieval paradigms. For English queries, dense retrieval achieves the strongest performance among all evaluated methods, with a Hit@10 of 0.5092, MRR@10 of 0.3501, and NDCG@10 of 0.3879. In comparison, BM25 obtains a Hit@10 of only 0.0743, with corresponding MRR@10 and NDCG@10 scores of 0.0343 and 0.0436. These results indicate that lexical overlap alone is insufficient for retrieving relevant jobs from English multi-turn conversations, whereas multilingual dense representations capture substantially more of the semantic correspondence between conversational requirements and job descriptions.

\paragraph{\textbf{Retrieval behavior changes substantially for Banglish.}}
The relative behavior of the two retrieval paradigms changes under Banglish input. Dense retrieval remains competitive, achieving a Hit@10 of 0.3558, MRR@10 of 0.2193, and NDCG@10 of 0.2517, although its effectiveness decreases relative to English. In contrast, BM25 improves markedly from a Hit@10 of 0.0743 on English to 0.3211 on Banglish. Although dense retrieval still outperforms BM25 under the Banglish condition, the substantially different response of the two retrieval mechanisms to linguistic variation suggests that lexical and semantic evidence provide complementary signals. We examine this behavior further under RQ3.

\paragraph{\textbf{Hard filtering consistently reduces retrieval effectiveness.}}
The hard-constraint variants underperform their corresponding unfiltered retrievers in both language conditions. For dense retrieval, Hit@10 decreases from 0.5092 to 0.2892 on English and from 0.3558 to 0.1947 on Banglish. Similar degradation is observed for sparse retrieval: BM25 decreases from 0.0743 to 0.0458 on English and from 0.3211 to 0.2251 on Banglish. The same pattern is reflected in MRR@10 and NDCG@10. Because the English and Banglish variants of each dialogue share the same structured preference annotations and therefore the same filtered candidate set, these results indicate that the loss is introduced by candidate elimination rather than by differences in the underlying recommendation target.

\paragraph{\textbf{Binary compatibility decisions can irreversibly remove relevant jobs.}}
Even when a preference is available, strict compatibility tests can reject postings that are plausibly relevant. For example, a conversational preference such as ``degree in accounting'' may not produce a direct textual match with a structured requirement represented as ``Bachelor,'' despite their potential compatibility. Similar discrepancies may arise from location normalization and numerical experience ranges. Because hard filtering is applied before retrieval, exclusion of the ground-truth posting imposes an upper bound on subsequent retrieval effectiveness: once the relevant posting has been removed from the candidate set, neither sparse nor dense ranking can recover it. The consistent degradation observed in Table~\ref{tab:model_performance} therefore supports the use of graded compatibility scores rather than irreversible binary exclusion.

\paragraph{\textbf{W-SCAR substantially improves over hard-constraint ranking.}}
The results for RQ2 demonstrate the benefit of replacing binary candidate exclusion with graded compatibility. W-SCAR achieves Hit@10 values of 0.3737 on English and 0.3843 on Banglish, compared with 0.2892 and 0.1947, respectively, for Filter+Dense. The improvement is particularly pronounced for Banglish, where the absolute Hit@10 gain over Filter+Dense is 0.1896. W-SCAR similarly exceeds Filter+BM25 under both language conditions and provides corresponding improvements in MRR@10 and NDCG@10. These results indicate that incorporating structured preferences as graded ranking criteria preserves useful constraint information without incurring the recall loss associated with binary filtering.

\paragraph{\textbf{W-SCAR provides the strongest performance on Banglish.}}
The advantage of the proposed formulation becomes more apparent under Banglish queries. W-SCAR achieves the highest performance across all three evaluation metrics, with a Hit@10 of 0.3843, MRR@10 of 0.2340, and NDCG@10 of 0.2694. The corresponding values for dense retrieval are 0.3558, 0.2193, and 0.2517, while BM25 obtains 0.3211, 0.1262, and 0.1717. Thus, combining lexical relevance, multilingual semantic similarity, and graded constraint compatibility is particularly beneficial when the conversational language differs from the predominantly English representation of the job corpus.

\paragraph{\textbf{Soft-constraint ranking yields substantially more balanced cross-lingual performance.}}
W-SCAR also exhibits considerably smaller variation between the two linguistic conditions than either sparse or dense retrieval alone. Its Hit@10 changes by only 0.0106 between English and Banglish, while MRR@10 changes from 0.2406 to 0.2340 and NDCG@10 from 0.2718 to 0.2694. In contrast, dense retrieval decreases by 0.1534 in Hit@10 from English to Banglish, while BM25 changes by 0.2468 in the opposite direction. These results suggest that the complementary signals incorporated by W-SCAR reduce dependence on any single linguistic representation.

\paragraph{\textbf{Sparse retrieval benefits from lexical cues in Romanized Banglish.}}
BM25 exhibits increased Hit@10 on Banglish (0.3211) compared to English (0.0743). This behavior is consistent with the presence of discriminative Latin-script terms, English technical expressions, local abbreviations, and code-mixed lexical cues in the Romanized conversations. Such terms can provide direct lexical matches with the predominantly English job corpus even when the surrounding dialogue contains Romanized Bangla. 

\paragraph{\textbf{Dense similarity scores provide limited separation among top-ranked candidates.}}
Inspection of the dense-retrieval scores provides an additional diagnostic. Across the unfiltered runs, the top-ten cosine similarities occupy relatively narrow ranges: approximately $[0.877, 0.922]$ for English and $[0.878, 0.915]$ for Banglish. Within individual queries, the mean difference between the first- and tenth-ranked candidates is only 0.018 for English and 0.014 for Banglish. These concentrated scores indicate that many candidate jobs occupy closely neighboring regions of the multilingual embedding space. Consequently, relatively small changes in query representation may alter the ordering of highly similar candidates, which may contribute to the sensitivity of dense retrieval across linguistic variants.

\paragraph{\textbf{Lexical and semantic retrieval provide complementary cross-lingual evidence.}}
Taken together, the results reveal complementary behavior rather than a uniformly superior retrieval mechanism. Dense retrieval provides strong semantic matching for English and remains effective for Banglish, whereas BM25 benefits substantially from lexical anchors in Romanized code-mixed queries. W-SCAR exploits both signals while representing structured preferences as graded utilities, producing Hit@10 values of 0.3737 and 0.3843 on English and Banglish, respectively. The small difference between the two language conditions supports the central motivation of the proposed framework: conversational job recommendation in linguistically heterogeneous settings benefits from combining semantic relevance, lexical evidence, and soft constraint compatibility rather than relying on a single retrieval signal or strict candidate filtering.

\section{Conclusion}
We introduced W-SCAR, a soft-constraint ranking framework for conversational job recommendation, and evaluated it on JobCCC, comprising 22,410 job postings and 988 aligned English and Banglish dialogues. Across both language conditions, hard-constraint filtering consistently reduced retrieval effectiveness, showing that binary eligibility checks can remove relevant jobs before ranking. W-SCAR avoids this failure mode by combining lexical relevance, dense semantic similarity, and graded compatibility for experience, location, education, and salary.

The results also reveal a clear cross-lingual tradeoff. Dense retrieval performs best on English, whereas W-SCAR achieves the strongest performance on Banglish and substantially reduces the performance gap between the two language conditions. These findings indicate that soft constraint integration is particularly useful when conversational preferences are noisy, incomplete, or expressed in code-mixed language.

\bibliographystyle{ACM-Reference-Format}
\bibliography{sample}

\appendix
\end{document}